\documentstyle[epsf,epsfig,twoside,longtable,12pt]{article}

\begin{document}

\vspace*{-1.8cm}
\hspace*{12cm}{\bf LAL 99-01}\\
\vspace*{-0.5cm}
\hspace*{12cm}{January 1999}
\vskip 7.5 cm

\begin{center}
{\bf\LARGE $B^0$ - $\bar B^0$ oscillations and measurements of\\
\vskip 2mm
$\vert V_{ub}\vert/\vert V_{cb}\vert$ at LEP}
\end{center}

\vskip 3 cm
\begin{center}
{\bf\Large  Achille Stocchi}
\end{center}

\vspace*{0.5cm}

\begin{center}
{\bf\large Laboratoire de l'Acc\'el\'erateur Lin\'eaire}\\

{IN2P3-CNRS et Universit\'e de Paris-Sud, BP 34, F-91898 Orsay Cedex}
\end{center}

\vskip 1 cm
\begin{center}
{\bf\Large }
\end{center}

\vspace*{-1cm}

\begin{center}
{\bf\large }
\end{center}

\vskip 0.5 cm
\begin{center}
{\it Talk given at the ``HQ98 Conference''\\
Fermilab - Batavia (USA), October 10-12, 1998}
\end{center}
\newpage

\title{$\rm{B}^0 - \overline{\rm{B}}^0$ oscillations and measurements of
$|\rm{V}_{ub}|/|\rm{V}_{cb}|$ at LEP}
\author{Achille Stocchi}
\maketitle
\begin{center}
Laboratoire de l'Acc\'el\'erateur Lin\'eaire\\
IN2P3-CNRS et Universit\'e de Paris-Sud\\
B.P. 34 - 91898 Orsay Cedex
\end{center}

\begin{abstract}
In this paper a review of the LEP analyses on  $\rm{B}^0-\overline{\rm{B}}^0$
oscillations and on the measurement of $|\rm{V}_{ub}|/|\rm{V}_{cb}|$
is presented .
These measurements are of fundamental importance in constraining the
$\rho$ and $\eta$ parameters 
of the CKM 
matrix. A review of the current status of the $\rm{V}_{\rm{CKM}}$ 
matrix determination is also 
given.
\end{abstract}
\section*{Introduction}
The data registration at the Z$^0$ pole has stopped at the end of 1995.
The four LEP experiments (ALEPH, DELPHI, L3 and OPAL) have collected about
4M hadronic Z$^0$ decays per experiment.

In the past three years, the quality of the data analysis has continuously 
improved,
thanks to a better understanding of the behaviour of all components 
of the detector.
At the same time, new ideas, and then, new analyses have been tried. A
more performant statistical treatment of the information has been
also developed. As a result, the precision on the $\Delta
\rm{m}_d$ parameter has been improved and above all, the sensitivity for 
the $\Delta \rm{m}_s$ parameter has been tremendously increased.
The new and precise LEP analyses on $|\rm{V}_{ub}|/|\rm{V}_{cb}|$ are also a 
consequence of these improvements. Many analyses described in this paper
have been presented at the last '98 Summer Conferences and are still preliminary. 
This paper is organized as 
follows. The first sections are dedicated to the 
oscillations and $|\rm{V}_{ub}|/|\rm{V}_{cb}|$
analyses. In the last section the present status of the $\rm{V}_{\rm{CKM}}$
matrix is given with a special emphasis placed on 
the impact of the measurements
presented in this paper.

\section*{The oscillation analyses}
The probability that a $\rm{B}^0$ meson oscillates into a 
$\overline{\rm{B}}^0$ or
stays as a $\rm{B}^0$ is given by:
\begin{equation}
P_{\rm{B}^0_q \rightarrow \rm{B}^0_q(\overline{\rm{B}}^0_q)} =
\frac{1}{2}e^{-t/\tau_q} (1 \pm cos \Delta \rm{m}_q t)
\end{equation}
where the effect of CP violation has been neglected. $\tau_q$ is the
lifetime of the $\rm{B}^0_q$ meson, $\Delta \rm{m}_q = \rm{m}_{B^0_1}
- m_{\rm{B}^0_2}$ is the mass difference between the two mass eigenstates\footnote
{$\Delta \rm{m}_q$ 
is usually
given in ps$^{-1}$. 1 ps$^{-1}$ corresponds to 6.58 10$^{-4}$eV.} and
gives the period of the time oscillations (the effect of a lifetime difference
between the two states has been also neglected).

The Standard Model predicts:
\begin{equation}
\Delta \rm{m}_d  ~\propto  ~A^2 \lambda^6 [(1 - \rho)^2 + \eta^2]
f^2_{\rm{B}_d} \rm{B}_d~;~ \Delta \rm{m}_s  ~\propto  ~A^2 \lambda^4 
f^2_{\rm{B}_s} \rm{B}_s
\end{equation}
The difference in the $\lambda$ dependence of these expressions ($\lambda \sim
0.22$) implies
that
$\Delta \rm{m}_s \sim 20 ~\Delta \rm{m}_d$. It is then 
clear that a very good proper
time resolution is needed to measure the $\Delta \rm{m}_s$ parameter.
A time dependent study of $\rm{B}^0 - 
\overline{\rm{B}}^0$ oscillations requires: \\
- the measurement of the decay proper time, \\
- to know if a $\rm{B}^0$ or a  $\overline{\rm{B}}^0$ decays at t = t$_o$
(decay tag)\\
- to know if a b or a $\overline{\rm{b}}$ quark has been produced 
at t = 0 (production
tag). 
\vspace{1.5cm}
\begin{figure}[h]
\begin{center}
\epsfig{file=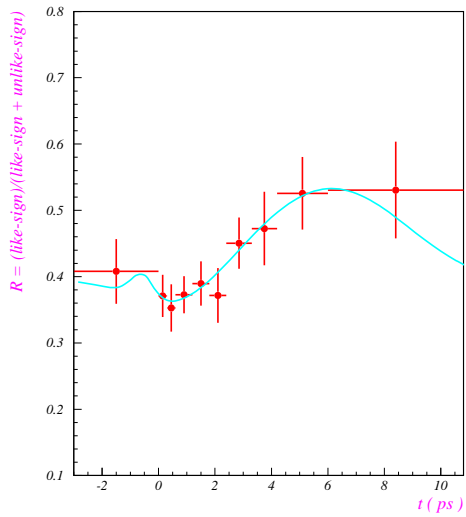,width=7cm}
\vspace{-1.5cm}
\caption{\small The plot shows the time dependence behaviour of 
the $\rm{B}^0_d - \overline{\rm{B}}^0_d$
oscillation. The points with error bars are the data. The curve shows
the
result of the fit using $\Delta \rm{m}_d = 0.47$ ps$^{-1}$.}
\label{figa}
\end{center}
\end{figure}

The precision on the $\Delta \rm{m}$ measurement is given by the following
relation:
\begin{equation}
\rm{error}~~= \left(\sqrt N f_{\rm{B}^0_{d(s)}} (2 \varepsilon_1 - 1)
(2 \varepsilon_2 - 1) e^{-\left(\frac{\Delta \rm{m}_{d(s)} \sigma_t}{2}
\right)^2}\right)^{-1}
\end{equation}
where N is the total number of events in the sample; $f_{\rm{B}^0_{d(s)}}$ 
is the fraction of events in which a $\rm{B}^0_{d(s)}$ 
meson has been produced; $\varepsilon_2, \varepsilon_1$ are the 
tagging purities at the decay
and production times respectively, defined as $\varepsilon =
\frac{N_{\rm{right}}}{N_{\rm{right}} + N_{\rm{wrong}}}$, where $N_{\rm{right}}
(N_{\rm{wrong}})$ are the numbers of correctly (incorrectly) tagged events 
and $\sigma_t$ is the proper time resolution given, approximately, as 
$\sigma_t =
\sqrt{\left( \frac{m^2}{p^2}\right) \sigma^2_L + \left( \frac{\sigma_p}
{p} \right)^2 t^2}$, where $\sigma_L$ and $\sigma_p$ are the decay length
and the momentum resolutions respectively. 

\subsection*{$\Delta \rm{m}_d$ measurements}

\begin{figure}[h]
\begin{center}
\epsfig{file=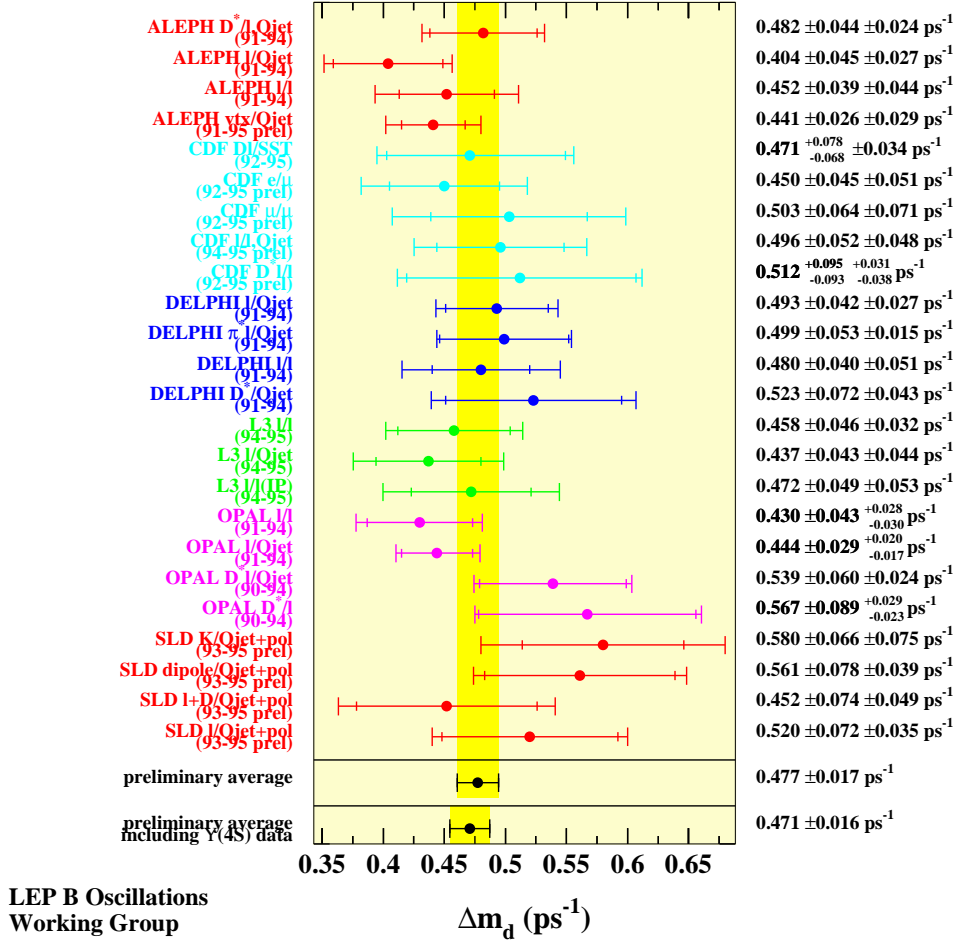,width=14cm}
\vspace{1cm}
\caption{\small Summary of the $\Delta \rm{m}_d$ results from the LEP,
SLD and CDF Collaborations are given. Details on how the 
different results have been
combined are given in \cite{ref1}.}
\label{fig1}
\end{center}
\end{figure}

A lot of analyses have been performed since 1994. A typical time distribution
is shown
in Figure \ref{figa}. The time dependence behaviour with frequency $\Delta \rm{m}_d$
$\sim 0.470~\rm{ps}^{-1}$, for the $\rm{B}^0_d - \overline{\rm{B}}^0_d$ 
oscillation
is clearly visible. This will be a textbook plot ! The present summary 
of the results on
$\Delta \rm{m}_d$, as given by \cite{ref1}, is shown in Figure \ref{fig1}. Combining
LEP/CDF and SLD measurements it follows that:
\begin{equation}
\Delta \rm{m}_d = (0.477 \pm 0.017) \rm{ps}^{-1}
\end{equation}
$\Delta \rm{m}_d$ is known with a precision of 3.4\% relative error.

\subsection*{Analyses on $\Delta \rm{m}_s$}

Four types of analyses have been performed.

\begin{table}[h]
\caption{\small The characteristics of the different analyses are given 
in terms of statistics (N), 
$\rm{B}^0_s$ purity $(f_{\rm{B}_s})$, tagging purity at the production 
and decay time $(\varepsilon_1, \varepsilon_2)$ and time resolution
in the first pico-second}
\begin{center}
\begin{tabular}{cccccc} \hline
Analysis & N(events) & $f_{\rm{B}_S}$ & $\varepsilon_1$ & $\varepsilon_2$ & $\sigma_t (t < 1 \rm{ps})$ \\ \hline
Inclusive lepton & $\sim 50000$ & $\sim 10\%$ & $\sim 70\%$ & $\sim 90\%$ & $\sim 0.25$ ps\\
$\rm{D}^\pm_s h^\mp$ & $\sim3000$ & $\sim 15\%$ & $\sim 72\%$ & $\sim 90\%$ & $\sim 0.22$ ps\\
$\rm{D}^\pm_s \ell^\mp$ & $\sim 400$ & $\sim 60 \%$ & $\sim 78 \%$ & $\sim 90 \%$ & $\sim 0.18$ ps\\
Exclusive $\rm{B}^0_S$ & $\sim 25$ & $\sim 70 \%$ & $\sim 78 \%$ & $\sim 100 \%$ & $\sim 0.08$ ps\\
\hline
\end{tabular}
\end{center}
\label{tab1}
\end{table}

For all of them, the latest analyses make use of the
combined tag method for tagging a b or a $\overline{\rm{b}}$ 
at production time. At LEP, the produced b 
and $\overline{\rm{b}}$ quarks fragment independently
and the events can be divided in two separate hemispheres. If the measurement
of the proper time is performed in one of those (same hemisphere), the
other (opposite hemisphere) can be used to determine if a 
b or a $\overline{\rm{b}}$ quark was produced in that hemisphere. 
Several variables are considered in the opposite hemisphere:\\
$\bullet$ $\rm{Q} = \frac{\sum^n_{i=1} q_i ({\vec p}_i . {\vec e}_S)^{0.6}}
{\sum^n_{i=1} ({\vec p}_i .  {\vec e}_S)^{0.6}}$ the hemisphere charge,
defined as the charge of all (n) charged tracks ($\rm{q}_i$) present 
in the hemisphere,
weighted by their momentum ($p_i$) projected along the thrust axis
(${\vec e}_S)$ with a chosen value for the exponent (0.6),\\
$\bullet$  the hemisphere charge, considering only identified kaons,\\
$\bullet$ the charge of primary and secondary vertices,\\
$\bullet$ the presence of high $p_t$ leptons.

The use of these variables allow to have a tagging purity of the order
of 70\%.
\vspace{-1cm}

Tracks in the same hemisphere can be used also. This procedure is 
peculiarly clean if all the tracks from the $\rm{B}^0_s$ have been 
reconstructed
(as for $\rm{D}_s^\pm \ell^\mp$ and exclusive $\rm{B}_s$ analyses). 
In this case, 
tracks from the $\rm{B}^0_s$ decay can be removed and the others, 
coming from the 
primary vertex can be used.
The addition of informations from the same hemisphere 
allows to reach a tagging purity of 74\%. Finally the use of all these
informations on an event by event basis gives a purity of 78\%.

The tagging of a B or a $\overline{\rm{B}}$ meson at decay time depends 
on the specific analysis and
will be given in the following. Before describing the
different analyses, the method used to measure or put a limit on
$\Delta \rm{m}_s$ is briefly discussed.

\subsubsection*{The amplitude method}

The method used to measure or to put a limit on $\Delta \rm{m}_s$ consists
in modifying equation 1 in the following way:
$1 \pm cos \Delta \rm{m}_s t \rightarrow 1 \pm A cos \Delta \rm{m}_s t$.
A and $\sigma_A$ are measured at fixed values  of
$\Delta \rm{m}_s$ instead of $\Delta \rm{m}_s$ itself. In case of
a clear oscillation signal, at given $\Delta \rm{m}_s$, the value
of the amplitude is compatible with A = 1 for this 
$\Delta \rm{m}_s$ and with A = 0 elsewhere.
With this method it is also easy to set a limit. The values of $\Delta
\rm{m}_s$ excluded at 95\% C.L. are those satisfying the condition A($\Delta
\rm{m}_s$) + 1.645 $\sigma_A (\Delta \rm{m}_s) < 1$.

With this method, it is easy to combine different experiments and to
treat systematic uncertainties in an usual way since, at each value of $\Delta
\rm{m}_s$, a value for A with a gaussian error $\sigma_A$, is measured. 
Furthermore,
the sensitivity of the experiment can be defined as the value of
$\Delta \rm{m}_s$ corresponding to 1.645 $\sigma_A (\Delta \rm{m}_s) = 1$ (for
A($\Delta \rm{m}_s) = 0$, namely supposing that the ``true'' value of
$\Delta \rm{m}_s$ is well above the measurable value of $\Delta \rm{m}_s$). The
sensitivity is the limit which would be reached in 50\% of the experiments.

\subsubsection*{The inclusive lepton/combined tag analysis}

This analysis uses high $p_t$ leptons which are mainly coming from
direct b semileptonic decays ($b \rightarrow \ell$). The sign of the lepton tags the
$\rm{B}^0_s$ at decay time. The initial sample consists in
80\% leptons from B decays (and among those 90\% $b \rightarrow \ell$ (direct) and 10\% 
$b \rightarrow c \rightarrow \overline{\ell}$ (cascade)) and of
20\% leptons from charm decays or misidentification.
The events $b \rightarrow c \rightarrow \overline{\ell}$ give the wrong tag
for the $\overline{\rm{B}}^0_s$ meson at decay time.

To reconstruct a B decay proper time, algorithms have
been developed which aim at identifying charged (neutral) tracks
which are more likely to come from the $\rm{B}^0_s$ decays. As result,
in more than 50\% of the cases, the error on the decay length
is $\sigma_L \sim 250 \mu m$ and the relative error on the B energy
is better than 10\%, resulting in an error on the proper time of the order
of 0.25 ps in the first pico-second.

A second crucial point for this analysis consists in trying to increase
the $\rm{B}^0_s$ purity of the sample (the natural $\rm{B}^0_s$ purity of b
events is around 10\%) and to reduce the contribution from cascade decays.
To enrich the sample in direct b semileptonic decays and, among those, in events
coming from $\rm{B}^0_s$ decays, several variables have been used as 
the momentum and transverse momentum of the lepton, the impact 
parameters of all tracks in the opposite hemisphere relative
to the main event vertex, the kaons in primary and 
secondary vertices in 
the same hemisphere, and the charge of the secondary vertex.

The result of this procedure is to increase the $\rm{B}^0_s$ purity by
30\% and to reach more than 90\% purity for the tagging at the decay
time. 

\subsubsection*{$\rm{D}_s^\pm \ell^\mp$/combined tag analysis}

The use of events in which a reconstructed $\rm{D}_s$ is accompanied by a high
$\rm{p}_t$ lepton with an electric charge opposite in sign allows to select
a sample having 60\% $\rm{B}_s$ purity. The proper time resolution benefits
also from the fact that the only missing particle is the
neutrino:
$\overline{\rm{B}}^0_s \rightarrow \rm{D}^+_s e^- \overline{\nu}_e$.
In the first pico-second the time resolution is about 0.18 ps in more than 80\%
of the events.

\begin{figure}[h]
\vspace{-0.5cm}
\begin{center}
\epsfig{file=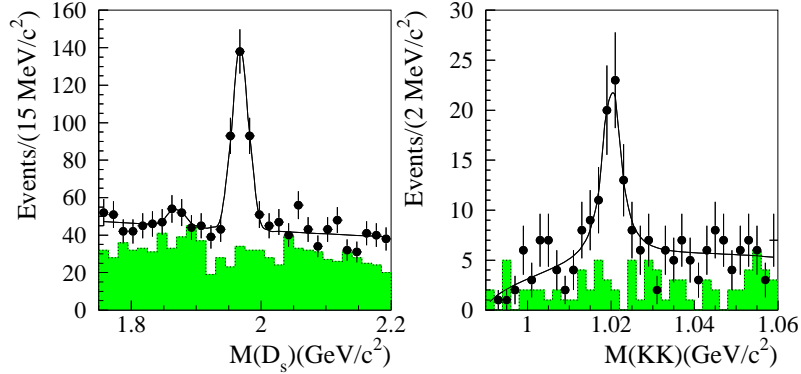,width=10.5cm}
\caption{\small DELPHI D$^\pm_s \ell^\mp$ candidates. The figure on the left
shows the ${\mathrm D}_s$ mass spectrum reconstructed from the following decay 
modes: $\mathrm{D}^+_s \rightarrow \phi \pi^+, \phi \pi^+\pi^0, \phi \pi^+
\pi^- \pi^+,$ $\overline{K}^{*0} K^+, \overline{K}^{*0} K^{+}
 \ \mathrm{and}\ K^0_S K^+$. The figure on the right shows the $\phi$ mass 
spectrum from the decays $\rm{D}^+_s \rightarrow \phi e^+ \nu_e\ \rm{and}\
\phi \mu^+ \overline{\nu}_\mu$. The sum of the two samples gives 230 $\pm$ 18 
$\rm{B}^0_s$ candidates.}
\label{fig3}
\end{center}
\end{figure}

The limiting factor is the available statistics because
accessible $\rm{D}_s$ branching fractions are quite small 
(between $\sim$ 1\% and $\sim$ 5\%). 
Several decay modes have to be selected. Figure \ref{fig3} shows an example
in which six hadronic and two semileptonic decay modes have been reconstructed.

\subsubsection*{Exclusive $\rm{B}^0_s$/combined tag analysis}

\begin{figure}[h]
\begin{center}
\epsfig{file=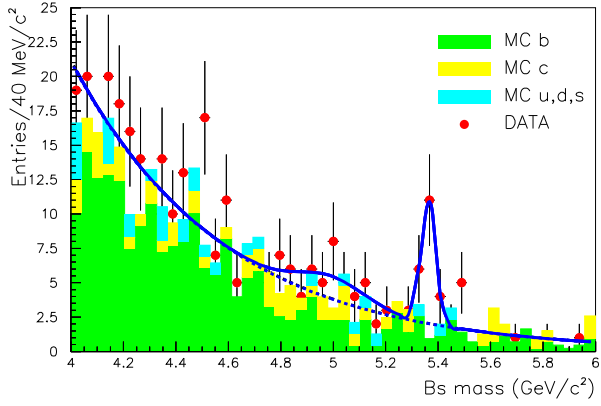,width=8cm}
\vspace{-3.5cm}
\caption{\small The $\rm{B}^0_s$ mass spectrum obtained by the DELPHI Collaboration.
The points with the error bars are the data with the fit superimposed. The
contributions from non-$\rm{B}^0_s$ decays, as given by the Monte Carlo
simulation, are also shown.}
\label{fig4}
\end{center}
\end{figure}

At the 1998 Moriond Conference, the DELPHI Collaboration has proposed the use
of exclusively reconstructed $\rm{B}^0_s$ decays for $\Delta \rm{m}_s$ analyses.
These events have an excellent proper time resolution $\sigma_t \sim
0.08$ ps and provide a gain in sensitivity at high
values of $\Delta \rm{m}_s$ (equation 3). Figure \ref{fig4} 
shows the $\rm{B}^0_s$ mass spectrum
using the decay modes: $\rm{B}^0_s \rightarrow \rm{D}_s \pi (\rm{or}~a_1)$ and 
$\rm{B}^0_s \rightarrow \rm{D}^0 K \pi (\rm{or}~a_1)$. The $\rm{D}_s$ has 
been reconstructed
in six hadronic decay modes, as in the $\rm{D}^\pm_s \ell^\mp$ analysis, and
the $\rm{D}^0$ is observed using $K \pi$ and $ K \pi \pi \pi$ decay modes. 
17 $\pm$ 
8 events have
been reconstructed in the $\rm{B}^0_s$ mass region. The combinatorial 
background is estimated to be 35\%.

\subsubsection*{Summary of $\Delta \rm{m}_s$ analyses}

\begin{figure}[h]
\begin{center}
\epsfig{file=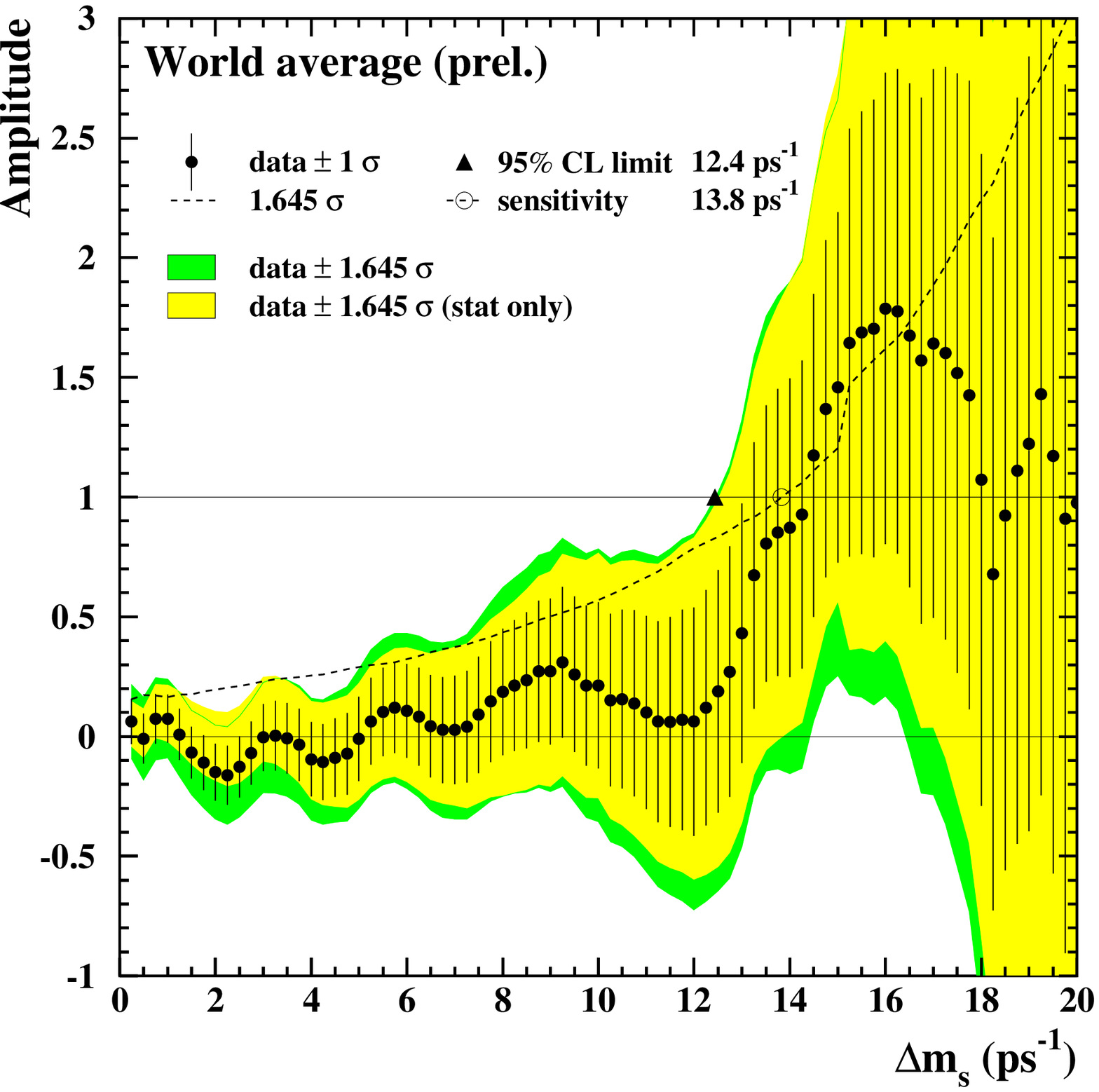,width=5.5cm,height=7.5cm}
\epsfig{file=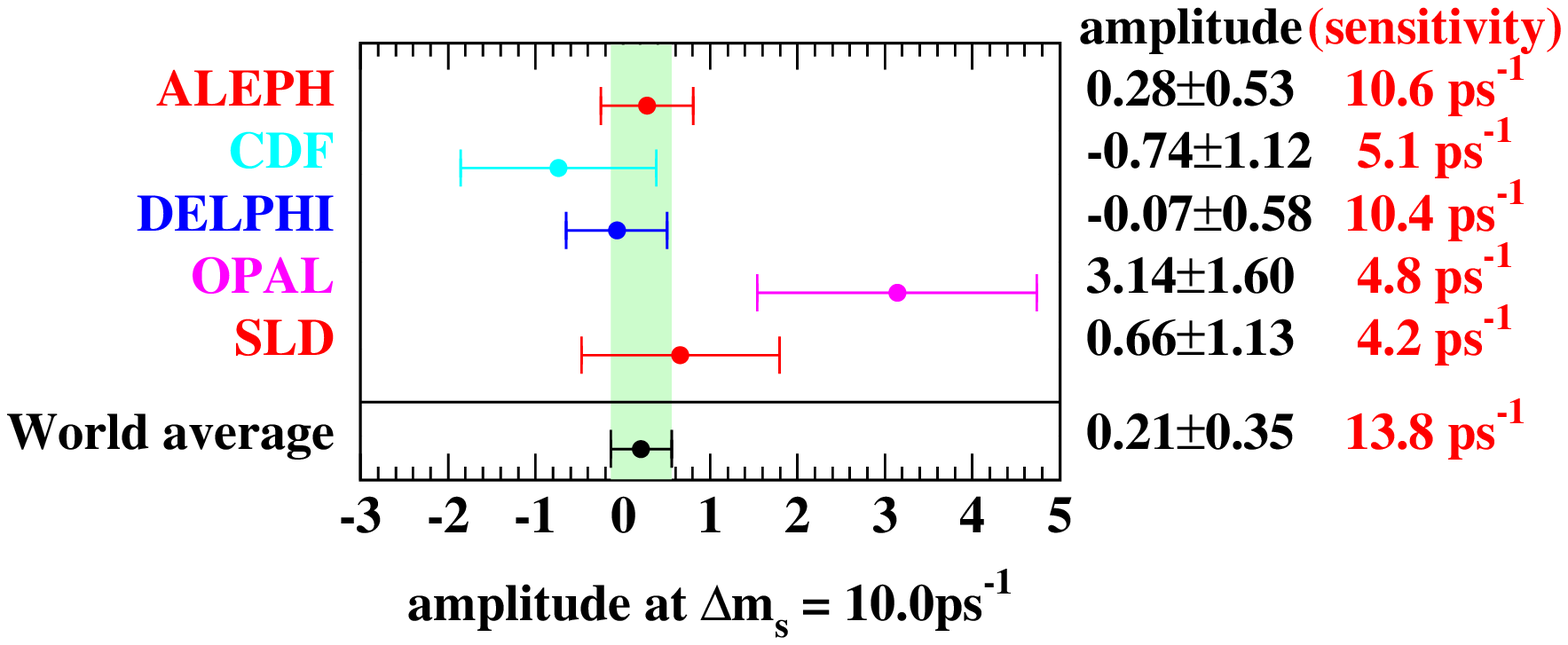,width=8cm,height=7.5cm}
\caption{\small The plot on the left shows the combined $\Delta \rm{m}_s$ 
results from LEP/SLD/CDF analyses
shown in an amplitude versus $\Delta \rm{m}_s$ plot. The point with 
error bars
are the data; the lines show the 95\% C.L. curves (in dark the
systematics have been included). The dotted curve shows the sensitivity.
The plot on the right shows the summary 
of the $\Delta \rm{m}_s$ results
per experiment. The
error are given at $\Delta \rm{m}_s = 10~\rm{ps}^{-1}$ 
(the sensitivity is also given). The way in which the combined 
value is obtained
is described in \cite{ref1}.}
\label{fig5}
\end{center}
\end{figure}

The combined result of LEP/SLD/CDF \cite{ref1} analyses is shown in
Figure \ref{fig5} and is:
$$\Delta \rm{m}_s > 12.4~\rm{ps}^{-1}~~\rm{at}~~95\%~~\rm{C.L.}$$
The sensitivity is at $13.8~\rm{ps}^{-1}$.
LEP alone has a limit at $11.5~\rm{ps}^{-1}$ at 95\% C.L., with a sensitivity
at $12.9~\rm{ps}^{-1}$. $\Delta \rm{m}_s$ = 0 is excluded between 
14.5~ps$^{-1}$ and $16.5~\rm{ps}^{-1}$
with a 2$\sigma$ significance at $15~\rm{ps}^{-1}$.
The present summary of the results is given in Figure \ref{fig5}.

\vspace{-2cm}

\clearpage
\section*{$|\rm{V}_{ub}|/|\rm{V}_{cb}|$ measurement}

The presence of leptons above the kinematical limit for those
produced in the decay $\rm{B} \rightarrow \rm{D} \ell \overline
{\nu}_\ell$ (b $\rightarrow$ c transition proportional to the $|\rm{V}_{cb}|$
CKM matrix element) is attributed to the transition $b \rightarrow
u \ell \overline{\nu}_\ell$ (proportional to the $|\rm{V}_{ub}|$ CKM matrix element).

The CLEO and ARGUS Collaborations have been pioneers 
in this measurement. Nevertheless, as only a small fraction of the energy
spectrum of these leptons is measurable, the systematic uncertainties
in the modelling of the b $\rightarrow$ u transition to evaluate the ratio 
$|\rm{V}_{ub}|/
|\rm{V}_{cb}|$
are quite large (of the order of 20\%-25\% relative error).
Recently LEP experiments have shown their capabilities 
of measuring $|\rm{V}_{ub}|$ with a statistical
precision similar to the one from CLEO and with reduced systematic
uncertainties. They use several kinematical variables,
in events with an identified high transverse momentum lepton, which
have a distinctive power to discriminate between b $\rightarrow$ c and 
b $\rightarrow$ u transitions. The first measurement has been 
performed by the ALEPH Collaboration by means of a neural network
discriminating method.

\begin{figure}[h]
\begin{center}
\epsfig{file=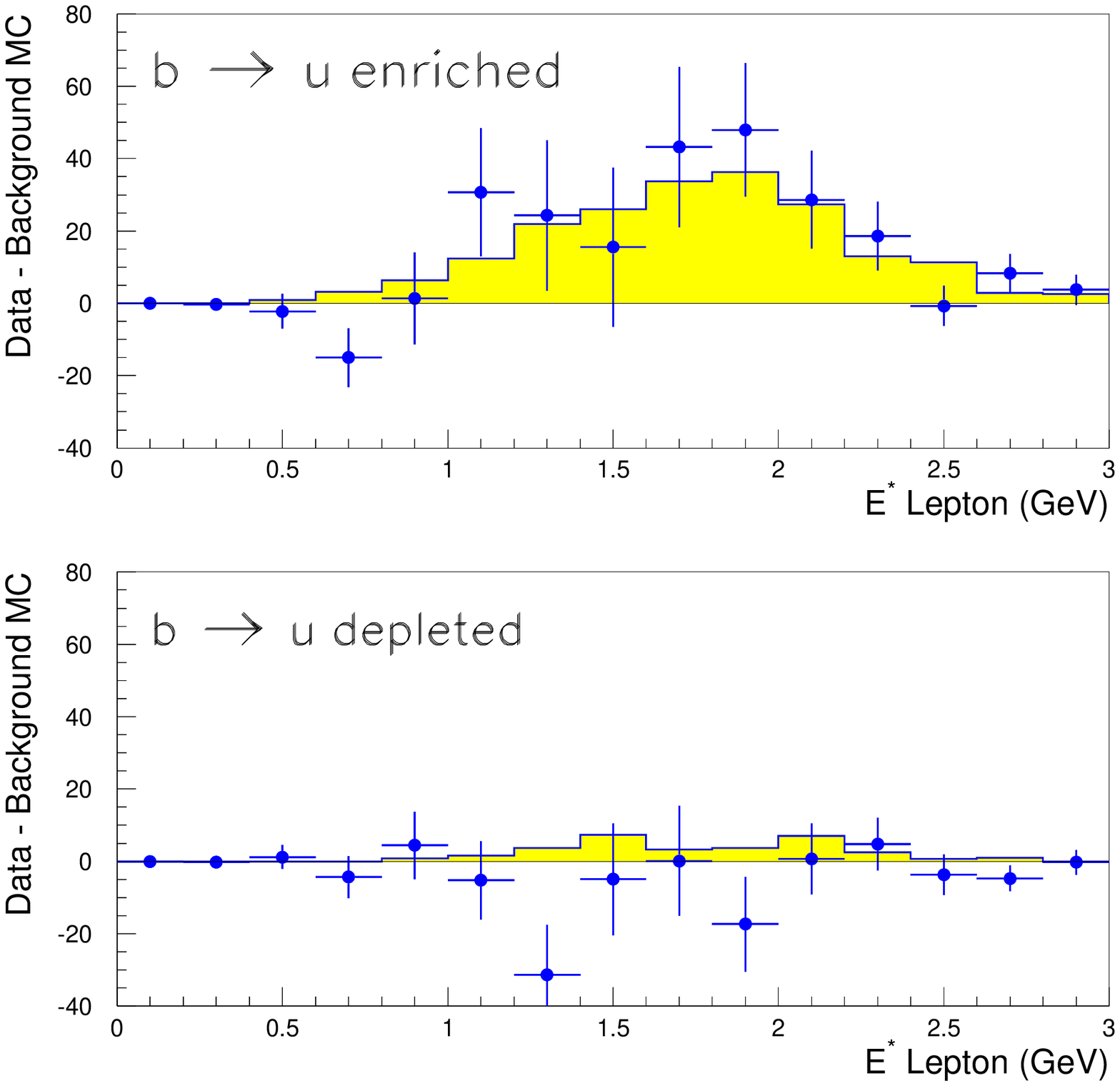,width=6.5cm}
\epsfig{file=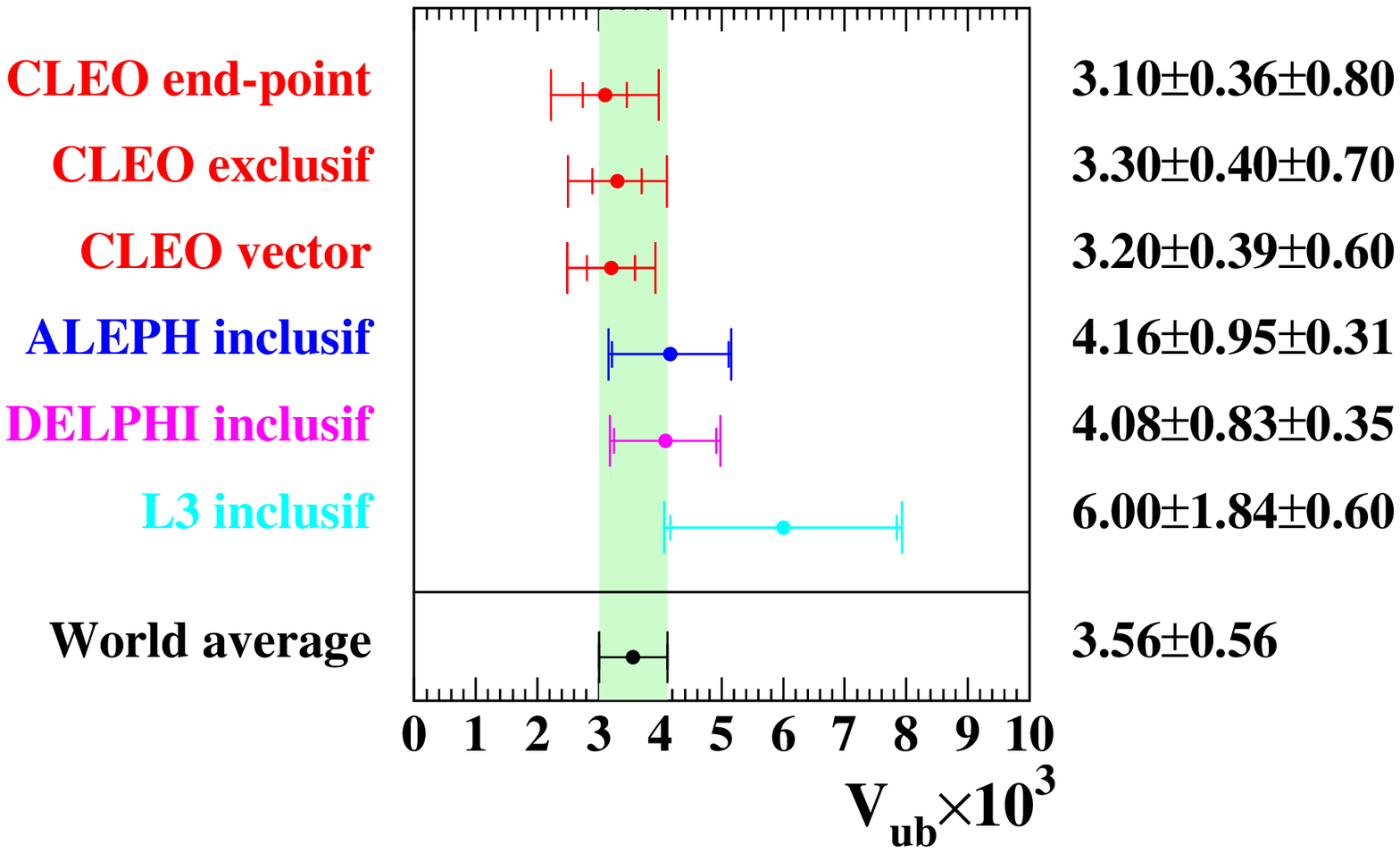,width=8.5cm}
\caption{\small The plots on the left show the energy of the lepton in
the B rest frame after the background subtraction for the $b \rightarrow
u$ enriched and $b \rightarrow u$ depleted samples. On the right the summary of
$|\rm{V}_{ub}|$ results is given.}
\label{fig6}
\end{center}
\end{figure}

The DELPHI measurement is simpler. With respect to the ALEPH analysis the
information from the presence of a secondary vertex from the D 
decay is used. In b $\rightarrow$ u transitions, all tracks are coming
from the B decay vertex. The presence of kaons at the D meson vertex
is also used. The method is based on the fact that the hadronic
system recoiling against the lepton in $b \rightarrow u \ell \nu$
decays is expected to have an invariant mass lower than the charm
mass \cite{ref2}. The sample is finally divided into a\linebreak 
b $\rightarrow$ u enriched
and a b $\rightarrow$ u depleted components and the energy of the lepton
in the B rest frame is calculated. The result is shown in Figure \ref{fig6} together
with the summary of the results on $|\rm{V}_{ub}|$.

\section*{Status of the $\rm{V}_{\rm{VCM}}$ matrix}

\begin{table}[h]
\caption{\small The four constraints which allow, at present, to define the accessible
region for the $\rho$ and $\eta$ parameters are listed in the first column.
In the second column the dependence of these constraints relative to the
different parameters is given. The last column gives
the explicit dependence in terms of $\overline{\rho}$
and $\overline{\eta}$.}
\vspace{0.5cm}
\begin{center}
\begin{tabular}{ccc}
\hline
Measurement & $\rm{V}_{\rm{CKM}} \times$ other & Constraint \\
\hline
b $\rightarrow$ u/ b $\rightarrow$ c & $(|\rm{V}_{ub}|/
|\rm{V}_{cb}|)^2$ & $\overline{\rho}^2 + \overline{\eta}^2$ \\
$\Delta \rm{m}_d$ & $|\rm{V}_{td}|^2 \rm{f}^2_{B_d}
\rm{B}_{B_d} \rm{f}(m_t)$ & (1 - $\overline{\rho})^2 +
\overline{\eta}^2$ \\
$\Delta \rm{m}_d/\Delta \rm{m}_s$ & $\left|\frac{\rm{V}
_{td}}{\rm{V}_{ts}}\right|^2  \frac{\rm{f}^2_{B_d} \rm{B}_{B_d}}
{\rm{f}^2_{B_d} \rm{B}_{B_s}}$ & (1 - $\overline{\rho})^2 +
\overline{\eta}^2$ \\
$\varepsilon_K$ & f(A, $\overline{\eta}, \overline{\rho},
\rm{B}_K)$ & $\sim \overline{\eta} (1 - \overline{\rho})$\\
\hline
\end{tabular}
\end{center}
\label{tab2}
\end{table}

The $\rm{V}_{\rm{CKM}}$ matrix can be parametrized in terms of four
parameters: $\lambda$, A, $\rho$ and $\eta$ (the Wolfenstein 
parametrization \cite{ref3}). The Standard Model predicts relations between
the different processes which depend on these parameters.
The unitarity of the $\rm{V}_{\rm{CKM}}$ matrix can be visualized
as a triangle in the $\rho - \eta$ plane.
Several quantities which depend on $\rho$ and $\eta$ have to be
measured and, if Standard Model is correct, they must define compatible
values for the two parameters inside measurement errors and theoretical
uncertainties. The measurement of b $\rightarrow$ u/ b $\rightarrow$ 
c transitions
gives a constraint of the form $\overline{\rho}^2 + \overline{\eta}^2$.\footnote{$\overline{\rho} (\overline{\eta}) = \rho(\eta) (1 - 
\lambda^2/2)$}
The measurement of $\Delta \rm{m}_d$  gives a constraint of the form (1 -
$\overline{\rho})^2 + \overline{\eta}^2$. A measurement of the ratio
$\Delta \rm{m}_d/\Delta \rm{m}_s$ gives the same type of constraint in
the $\overline{\rho} - \overline{\eta}$ plane, as a measurement of
$\Delta \rm{m}_d$, but this ratio is expected to have smaller
theoretical uncertainties since the ratio $\rm{f}^2_{B_s} \rm{B}_{B_s}/ 
\rm{f}^2_{B_d} \rm{B}_{B_d}$ is better known than the absolute value 
$\rm{f}^2_{B_d} \rm{B}_{B_d}$.

\begin{figure}[h]
\begin{center}
\epsfig{file=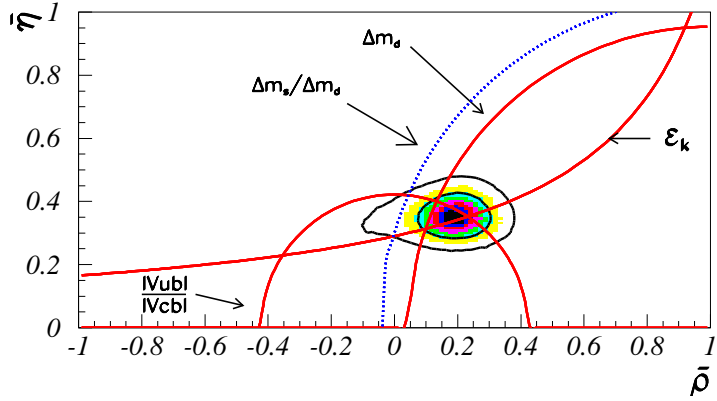,width=12cm}
\vspace{0.5cm}
\caption{\small The $\overline{\rho} - \overline{\eta}$ allowed region.
The contours at 68\% and 95\% C.L. are shown. The continous lines
correspond to the constraints obtained from the measurements of
$\frac{|\rm{V}_{ub}|}{|\rm{V}_{cb}|}, \Delta \rm{m}_d,$
and $\varepsilon_K$. The dotted curve corresponds to the 95\%
C.L. limit obtained from the experimental limit on $\Delta \rm{m}_s$.}
\label{fig8}
\end{center}
\end{figure}

All details of the analysis presented here can be found in \cite{ref4}.
Using the available and most recent measurements and up to date
theoretical calculations \cite{ref4} the allowed region in the $\overline{\rho}
- \overline{\eta}$ plane can be determined. It is shown in Figure \ref{fig8} 
and
corresponds to:
$$\overline{\rho} = 0.189 \pm 0.074~;~\overline{\eta} = 0.354 \pm 0.045$$

\begin{figure}[h]
\begin{center}
\epsfig{file=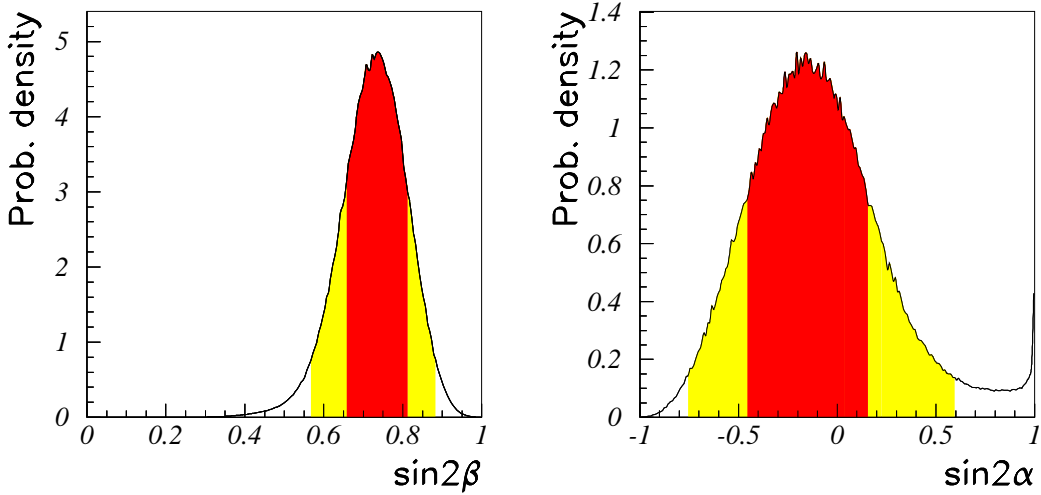,width=15cm}
\caption{\small The sin 2$\beta$ and sin 2$\alpha$ probability
density distributions. The dark-shaded and the clear
shaded intervals correspond to 68\% and 95\% C.L. regions 
respectively.}
\label{fig9}
\end{center}
\end{figure}

It is of interest to determine the central values and the uncertainties
on the quantities sin 2$\alpha$, sin 2$\beta$ and $\gamma$ which
will be directly measured at future B-factories or LHC
experiments. The result is 
shown in Figure \ref{fig9} and is:

\begin{equation}
\rm{sin} 2 \beta  =  0.73 \pm 0.08~;~\rm{sin} 2 \alpha  = - 0.15 \pm 0.30~;
~\gamma  =  (62 \pm 10)^0
\end{equation}

\begin{figure}[h]
\begin{center}
\epsfig{file=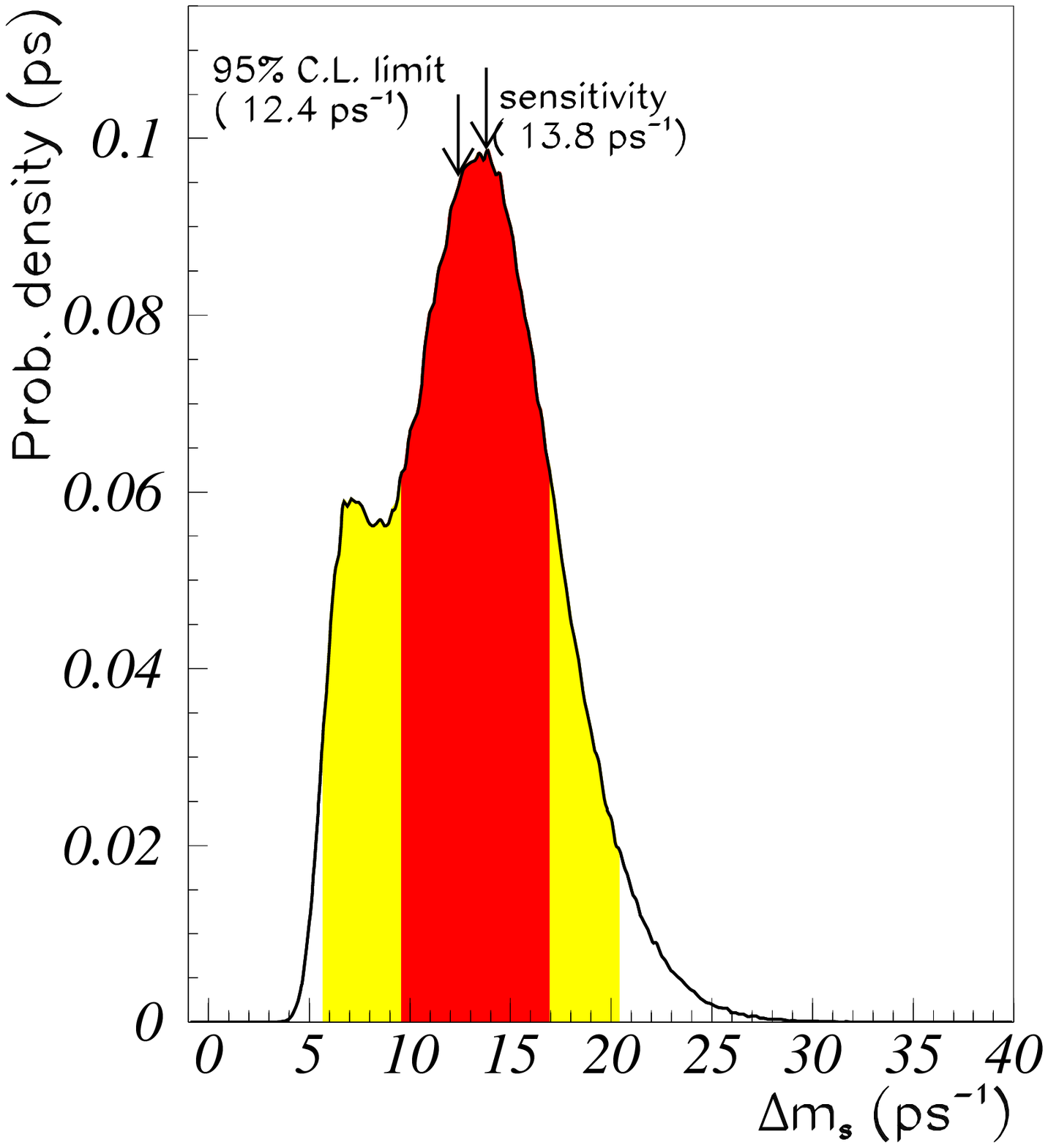,width=7cm}
\epsfig{file=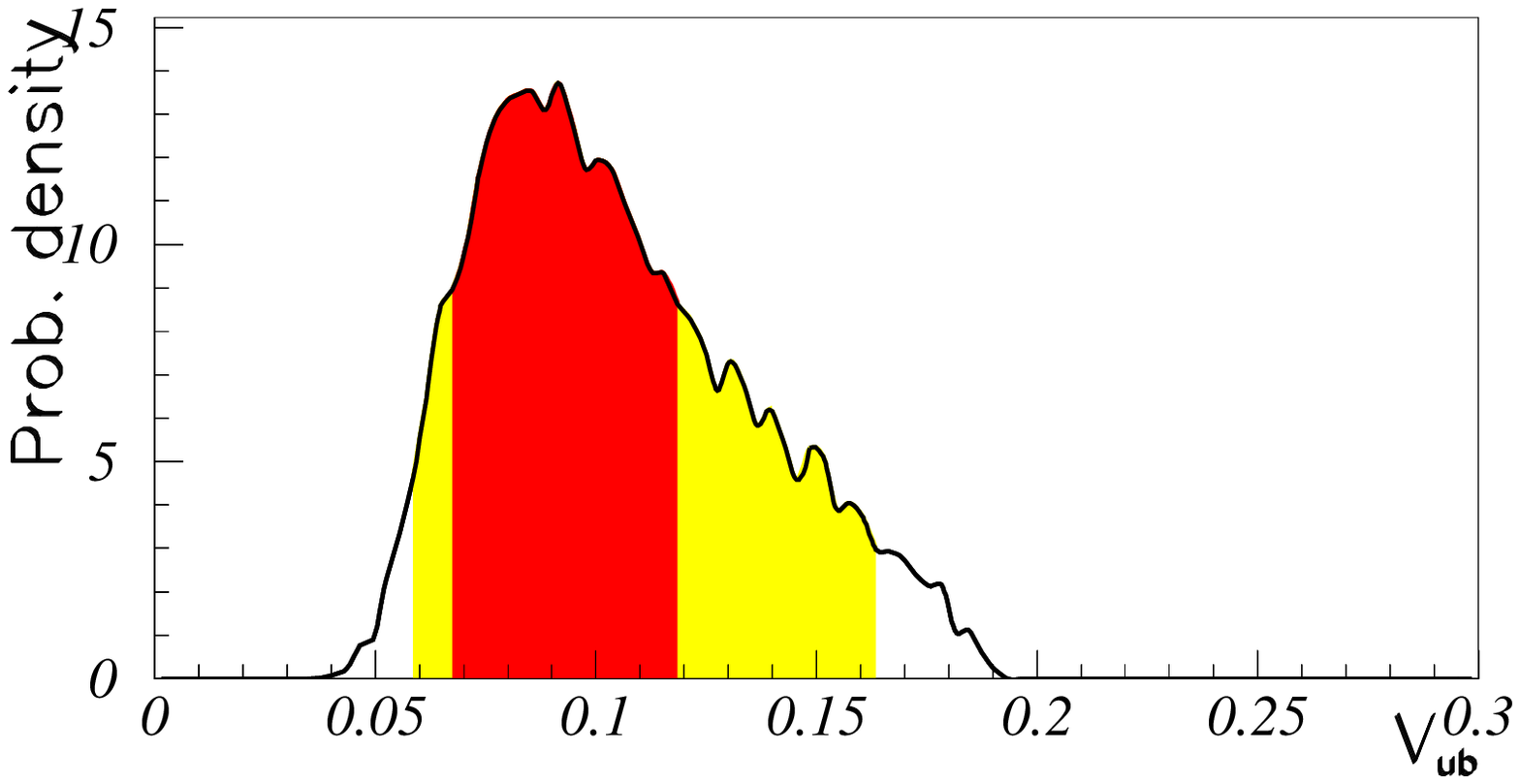,width=7cm}
\caption{\small The left and the right plots show the probability density
distributions for $\Delta \rm{m}_s$ and $|\rm{V}_{ub}|/
|\rm{V}_{cb}|$ respectively. The dark-shaded and the clear shaded intervals
correspond to 68\% and 95\% C.L. regions respectively.}
\label{fig10}
\end{center}
\end{figure}

The value of sin 2$\beta$ is rather precisely determined with an 
accuracy already at the level expected after the first years of 
running at B factories.
Finally it is possible to remove from the calculation the information
of one of the constraint and to obtain
its probability density
function. The result for $\Delta \rm{m}_s$ and $|\rm{V}_{ub}|/|\rm{V}
_{cb}|$ is shown in Figure \ref{fig10} and summarized in Table \ref{tab3}.

\clearpage
\def \textfraction{0}

\def \textfraction{0.2}
\clearpage

\begin{table}[h]
\caption{\small The $\Delta \rm{m}_s$ and $|\rm{V}_{ub}|/|\rm{V}
_{cb}|$ measured values are compared with those obtained using the fitting
procedure after having removed them from the fit.}
\begin{center}
\begin{tabular}{ccc}
\hline
Quantity & Measured value & Fitted value \\
\hline
$\Delta \rm{m}_s$ & $>$ 12.4~ps$^{-1}$ at 95\% C.L. & [9.5 - 17]~ps$^{-1}$ 68\% \rm{C.L.}\\
$|\rm{V}_{ub}|/|\rm{V}_{cb}|$ & 0.093 $\pm$ 0.014 & $0.085^{+0.037}_{-0.023}$\\
\hline
\end{tabular}
\end{center}
\label{tab3}
\end{table}
From these results the important impact of these two
measurements in the determination of the allowed region 
for $\rho$ and $\eta$ is clear. 
Furthermore the expected probability distribution 
for $\Delta \rm{m}_s$ shows that
present analyses are exploring the one sigma
region.

\section*{Conclusions}

Important improvements have been obtained in the last two years
in the analyses of $\rm{B}^0 - \overline{\rm{B}}^0$ oscillations.
Combining LEP results with those from
SLD and CDF, $\Delta \rm{m}_d$ frequency is presently known 
with a 3.4\% relative error
$(\Delta \rm{m}_d = 0.477 \pm 0.017~\rm{ps}^{-1})$.
The sensitivity on $\Delta \rm{m}_s$ is at 13.8~ps$^{-1}$ and, the 
actual LEP/SLD/CDF combined limit, of 12.4~ps$^{-1}$ at 95\% of C.L.,
is exploring the region where $\Delta \rm{m}_s$ is expected to be
according to the analysis \cite{ref4}. The measurement of $\Delta \rm{m}_s$ is
still a challenge for LEP collaborations, $|\rm{V}_{ub}|$ has
been measured at LEP with about the same experimental precision
as the one obtained by CLEO and with a reduced dependence on
theoretical models.

The phenomenological analysis presented in this paper gives:
$$\overline{\rho}  =  0.189 \pm 0.074~;~\overline{\eta} =  
0.354 \pm 0.045$$
and, in an indirect way:
$$sin2 \beta = 0.73 \pm 0.08~;~sin2 \alpha = -0.15 \pm 0.30~;~\gamma 
=  (62 \pm 10)^0$$
The situation will still be improved, at least until the
next summer '99, before the starting of B-factories.

\section*{Acknowledgement}

I would like to thank the organisers of HQ98 for the warm and nice
atmosphere during the conference and for the unforgettable banquet
at the Shedd Aquarium. Many thanks to Fabrizio Parodi and Patrick
Roudeau for their help in the preparation and redaction of this
contribution. Finally a grand merci to Jocelyne Brosselard, kind and
efficient as usual in the preparation of this manuscript.

\end{document}